\documentclass[a4paper,fleqn,usenatbib]{mnras}

\usepackage[T1]{fontenc}
\usepackage{ae,aecompl}

\usepackage{graphicx}	
\usepackage{amsmath}	
\usepackage{amssymb}	

\usepackage{amsmath,inputenx,nicefrac,mathrsfs}
\usepackage{xcolor}
\usepackage{graphicx}
\usepackage{multicol}
\usepackage{wasysym}
\usepackage{pstricks}
\usepackage{txfonts,slashbox}
\usepackage{nicefrac}
\usepackage{tikz,tkz-euclide}
\usetikzlibrary{quotes,angles}
\usetkzobj{all}
\usepackage{footnote,tablefootnote}
\makesavenoteenv{tabular}
\usepackage{cellspace}
\usepackage{multirow}
\usepackage{multicol}
\usepackage{cancel,empheq}

\title[On the local and global properties of the gravitational spheres of influences]{\textsc{On the local and global properties of the gravitational spheres of influence}}

\author[D. Souami et al.]{D. Souami$^{1,2}$\thanks{E-mail: damya.souami@obspm.fr ; souami@astro.utoronto.ca}
, J. Cresson$^3$, C. Biernacki$^4$, and F. Pierret$^5$%
\\
$^{1}$LESIA, Observatoire de Paris - Section Meudon, 5 Place Jules Janssen, 92195 Meudon Cedex.\\
$^{2}$naXys, University of Namur, Rempart de la Vierge, Namur, B-5000, Belgium. \\
$^{3}$Laboratoire de Math\'ematiques Appliqu\'ees de Pau, Universit\'e de Pau et des Pays de l'Adour, avenue de l'Universit\'e, BP 1155, 64013 Pau Cedex, France.\\
$^{4}$ Inria, Univ. Lille, CNRS, UMR 8524 - Laboratoire Paul Painlev\'e , F-59000 Lille. 40 av. Halley, 59650 Villeneuve d'Ascq, France.\\
$^{5}$IMCCE-CNRS UMR8028, Observatoire de Paris, PSL Universit\'e, Sorbonne Universit\'e, 77 Av. Denfert-Rochereau, 75014 Paris, France.\\ 
}

\date{Accepted 2020 May 22. Received 2020 May 16; in original form 2019 December 31.}

\pubyear{2020}

\begin{document}
\label{firstpage}
\pagerange{\pageref{firstpage}--\pageref{lastpage}}
\maketitle

\begin{abstract}
We revisit the concept of sphere of gravitational activity, to which we give both a geometrical and physical meaning. This study aims to refine this concept in a much broader context that could, for instance, be applied to exo-planetary problems (in a Galactic stellar disc-Star-Planets system) to define a first order "border" of a planetary system.\\     
   The methods used in this paper rely on classical Celestial Mechanics and develop the equations of motion in the framework of the 3-body problem (e.g. Star-Planet-Satellite System). We start with the basic definition of planet's \textit{sphere of activity} as the region of space in which it is feasible to assume a planet as the central body and the Sun as the perturbing body when computing perturbations of the satellite's motion. We then investigate the geometrical properties and physical meaning of the ratios of Solar accelerations (central and perturbing) and planetary accelerations (central and perturbing), and the boundaries they define.\\  
   We clearly distinguish throughout the paper between the sphere of activity, the Chebotarev sphere (a particular case of the sphere of activity), Laplace sphere, and the Hill sphere. The last two are often wrongfully thought to be one and the same. Furthermore, taking a closer look and comparing the ratio of the star's accelerations (central/perturbing) to that of the planetary acceleration (central/perturbing) as a function of the planeto-centric distance, we have identified different dynamical regimes which are presented in the semi-analytical analysis.
\end{abstract}




\section{Introduction}
\label{sect:introduction}
\vspace{-0.2cm}
The concept of \textit{spheres of influence} was first addressed by Pierre Simon \textsc{de Laplace} (1749 - 1827), in the context of the study of close encounters of comets with Jupiter.

 In his work entitled "Th\'eorie des Com\`etes", he investigated many aspects of the orbital dynamics of comets. In particular, he studied the motion of a comet which was about to pass in the vicinity of Jupiter in 1770 and evaluated the influence of Jupiter on the motion of the comet. He thus introduced the concept of the sphere of gravitational influence in Chapter II of \citep{Laplace1878}\footnote{We have chosen to cite the 1878 edition of \textsc{Laplace}'s work, as it is easily accessible via the French National Library online archives \url{https://gallica.bnf.fr/}.}, titled \textit{"Des perturbations du mouvement des com\`etes lorsqu'elles approchent tr\`es-pr\`es des plan\`etes"}.

This question was later studied by many authors in the context of \textit{the spheres of activity of the planets} \citep{1899PA......7..180H,1899PA......7..281M}, and is now a classical concept in every astrodynamics book \citep{1988ahl..book.....R}. The most commonly used models of spheres of influence are the Laplace Sphere \citep{1899PA......7..281M} and the Hill Sphere \citep{1878Hill}.

The necessity to revisit all these concepts of spheres of influence is related to two observations:
\vspace{-0.2cm}
\begin{itemize}
\item Firstly, many authors use the Hill and the Laplace Spheres interchangeably, as equivalent to one another and to the sphere of influence. However, the associated spheres have different physical meanings as shown throughout this paper. 
\item Secondly, in order to study the statistics of planetary distributions around a given star, one needs to introduce a cut-off radius on the possible zone where a planet no longer belongs to the given system. This cut-off distance is often chosen in an ad hoc way \citep{1998Icar..135..549H,1997A&A...322.1018N}. %
\end{itemize}

\cite{1984Natur.311...38S} looked into the problem of defining the outer border of the solar system by considering the 3-body system: Galactic centre -- Sun -- Oort cloud Comet. They performed numerical integrations looking into the comet's orbit around the Sun, concentrating on the shape of the boundary of the Solar System where the gravitational attraction of the Sun dominates over that of the rest of the Galaxy.

This question of delimiting the border of the solar system remains a complex open question with multiple associated answers. On page 77 of their book, \cite{Doressoundiram2008} give an overview of this question and a couple of associated answers. When dynamists look into the sphere of gravitational influence of the Sun as the volume where the Solar attraction dominates, plasma physicists would define the border as the region where solar winds are slowed down because of the interaction with the interstellar medium (at an estimated distance of 75 to 90 AU). 

This paper aims to provide a general framework in facilitating celestial mechanicians seeking to understand when, in an effective 3-body problem, one body's influence on a planetesimal (tiny mass) is replaced by the secondary body. 

By the use of analytical and semi-analytical methods, and in the framework of the 3-body problem (star, planet, and satellite), we revisit results concerning the definition of the limits of a planetary system by exploring the \textit{spheres of gravitational influence} of  the planet. We expand, in particular, on the work of G.A. Chebotarev \citep{1964SvA.....7..618C}.

Finally, in the appendix section, we apply these concepts to the Giant Planets and their systems of rings and satellites.

\section{Limits of a planetary system: Sphere of influence}
\label{sect:SOI}
Let us consider the following system: star (of mass $M$), planet (of mass $m$), and a planetesimal (of mass $m'$, with $m'<<m$), for example the Sun-Earth-Moon system. 

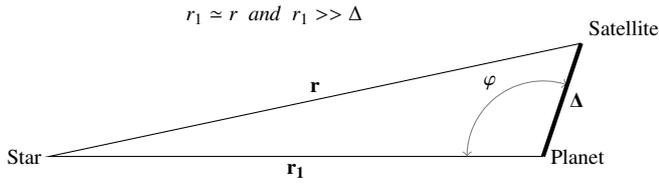
\begin{figure}
\begin{center}
\setlength{\unitlength}{0.9cm}
\begin{tikzpicture}
          \coordinate (A) at (6.5cm,0.cm);
          \coordinate (C) at (0.cm,0.cm);
          \coordinate (B) at (7.cm,1.5cm);
  \draw
    (6.5,-0.0) coordinate (a) node[right] {Planet} 
    -- (0,0) coordinate (b) node[left] {Star}
    -- (7,1.5) coordinate (c) node[above right] {Satellite}
    pic["$\varphi$", draw=gray, <->, angle eccentricity=1.2, angle radius=1cm]
    {angle=c--a--b};
\draw [ultra thick] (7,1.5) -- (6.5,-0.0);    
\tkzLabelSegment[below=0pt](C,A){$\mathbf{r_1}$}
\tkzLabelSegment[above=0pt](C,B){$\mathbf{r}$}
\tkzLabelSegment[ right=0pt](A,B){$\mathbf{\Delta}$}

 \put(2.0,2){ $r_1 \simeq r\,\,\, {and}\,\,\, r_1>> \Delta $}
\end{tikzpicture}
  \caption{\label{fig:3bp-rep} Geometry of the 3-body problem considered in this paper. We use bolded notations to indicate the vectors, and the non-bolded notations of these same quantities for the associated distances. $\mathbf{r_1}$ and $\mathbf{r}$ are the heliocentric position vectors of the planet and the satellite, respectively. 
  In the reference frame attached to the planet, the position vector of the satellite is $\mathbf{\Delta}=\mathbf{r}-\mathbf{r_1}$.
  }
\end{center}
\end{figure}

In all that follows, we will work in a heliocentric reference frame ($\mathscr{R}$). We denote by $\bf{r_1}$$(x_1,y_1,z_1)$ and $\bf{r}$$(x,y,z)$ the heliocentric position vectors of the planet and the satellite, respectively. $\Delta$~denotes the distance between the satellite and the planet, whereas $\varphi$ is the angle between the direction from the centre of the planet to the satellite, and to the Sun. The geometry of the problem is represented in Fig. \ref{fig:3bp-rep}.\\

\vspace{-0.5cm}

\subsection{The equations of motion}
In the framework of the three body problem for the system (star, planet, and satellite), G.A. Chebotarev wrote the equations of motion of the satellite (see \cite{1964SvA.....7..618C} for details). He considered the following two cases:
\vspace{-0.2cm}

\begin{itemize}
\item \textit{Case I.} The main body is considered to be the Sun, and the perturbing acceleration is caused by the attraction of the planet.
\end{itemize}
In the heliocentric reference frame ($\mathscr{R}$), we write the equations of motion:
\begin{itemize}
\item[] of the satellite 
\begin{equation}	
\mathbf{\ddot{r}}=-\frac{GM}{r^3}\mathbf{r} + Gm\bigg[\frac{\mathbf{r_1}-\mathbf{r}}{\Delta^3} - \frac{\mathbf{r_1}}{r_1^3}\bigg] \quad ,
\label{eqn:SatHelio}
\end{equation}
\item[] of the planet
\begin{equation}	
\mathbf{\ddot{r}_1}=-\frac{G(M+m)}{r_1^3}\mathbf{r_1} \quad .
\end{equation}
\end{itemize}		

In what follows, $G$ denotes the universal gravitational constant. Using the same notations as in \cite{1964SvA.....7..618C}, Equation \ref{eqn:SatHelio} writes as 
\begin{equation}	
\mathbf{\ddot{r}}=\mathbf{R}+\mathbf{F}\quad , \label{eqn:MainHelio}
\end{equation}
where $\mathbf{R}$ the acceleration which the central body (star) imparts to the satellite, and $\mathbf{F}$ the perturbing acceleration caused by the planet. These accelerations write as follows:
\begin{equation}	
 \left\{ \begin{array}{lcl}
       \mathbf{R}&= &-\frac{GM}{r^3}\mathbf{r} \\
       \mathbf{F}&= & Gm\bigg[\frac{\mathbf{r_1}-\mathbf{r}}{\Delta^3} - \frac{\mathbf{r_1}}{r_1^3}\bigg]\quad . \label{eqn:acc_helio}
\end{array} \right.
\end{equation}	

\begin{itemize}
\item \textit{Case II.} The planet is assumed to be the central body, and the perturbing acceleration is that imparted by the Sun.
\end{itemize}

In the reference frame ($\mathscr{R}_1$) attached to the planet, we write the position vector of the satellite $\mathbf{\Delta}=\mathbf{r}-\mathbf{r_1}$. The equations of motion in this reference frame write as follows:
\begin{equation}	
\mathbf{\ddot{\Delta}}=-\frac{Gm}{{r'}^3}\mathbf{\Delta} + GM\bigg[\frac{\mathbf{r_1}}{r_1^3} - \frac{\mathbf{r}}{r^3}\bigg] \quad . \label{eqn:SatGeo}
\end{equation} %
Similarly, using the same notations as in \cite{1964SvA.....7..618C}, Equation (\ref{eqn:SatGeo}) writes as follows 
\begin{equation}	
\mathbf{\ddot{\Delta}}=R_1+F_1\quad,
\end{equation}
where $\mathbf{R_1}$ is the acceleration which the central body (planet) imparts to the satellite, and $\mathbf{F_1}$ is the perturbing acceleration caused by the Star. These accelerations write as follows
\begin{equation}	
\left\{ \begin{array}{lcl}
       \mathbf{R_1}&= & -\frac{Gm}{{r'}^3}\mathbf{\Delta} \\
       \mathbf{F_1}&= & GM\bigg[\frac{\mathbf{r_1}}{r_1^3} - \frac{\mathbf{r}}{r^3}\bigg] \quad . \label{eqn:acc_geo}
\end{array} \right.
\end{equation}

\subsection{Developing the perturbing accelerations}
We develop the moduli of the accelerations ($F\,\text{and} \, F_1$) given by Equations (\ref{eqn:acc_helio}) and (\ref{eqn:acc_geo}) in terms of the Cartesian coordinates for the vectors $\mathbf{r}$$(x,y,z)$ and $\mathbf{r_1}$$(x_1,y_1,z_1)$. It follows that
\begin{equation}	
\left\{
\begin{array}{lll}
       R= \frac{GM}{r^2} \\
       F=  Gm\Bigg[\bigg(\frac{{x_1}-{x}}{\Delta^3} - \frac{{x_1}}{r_1^3}\bigg)^2 +\bigg(\frac{{y_1}-{y}}{\Delta^3} - \frac{{y_1}}{r_1^3}\bigg)^2 +\bigg(\frac{{z_1}-{z}}{\Delta^3} - \frac{{z_1}}{r_1^3}\bigg)^2	\Bigg]^{\nicefrac{1}{2}} \quad ,
       \label{eqn:acc_helio2}
\end{array} \right.
\end{equation}	
and
\begin{equation}	
\left\{
\begin{array}{lll}
       R_1 = \frac{Gm}{\Delta^2} \\
       F_1=  GM \Bigg[\bigg(\frac{x_1}{r_1^3} - \frac{{x}}{r^3}\bigg)^2 + \bigg(\frac{y_1}{r_1^3} - \frac{{y}}{r^3}\bigg)^2 + \bigg(\frac{z_1}{r_1^3} - \frac{{z}}{r^3}\bigg)^2  \Bigg]^{\nicefrac{1}{2}} \quad .
        \label{eqn:acc_geo2}
\end{array} \right.
\end{equation}	

For the sake of simplicity in the equations, we introduce the following notations (similar to \cite{1964SvA.....7..618C})
\begin{equation}	
\xi = x-x_1	\quad , \quad \eta=y-y_1 \quad , \quad \zeta = z-z_1\quad . \label{eqn:DeltaCoord}
\end{equation}

Furthermore, from the geometry of the three body problem (cf.~Fig.~\ref{fig:3bp-rep}), it follows that:
\begin{equation}	
\frac{x_1x+y_1y+z_1z}{r_1\Delta} = \cos\varphi \quad . \label{eqn:varphi}
\end{equation}	
 We introduce the following notation
\begin{equation}	
u=\frac{\Delta}{r_1} \label{eqn:u} \quad .
\end{equation}	

Using Equations (\ref{eqn:DeltaCoord}) and (\ref{eqn:varphi}), one can simplify the expressions of the perturbations $F$ and $F_1$ (given by Equations (\ref{eqn:acc_helio2}) and (\ref{eqn:acc_geo2}), respectively).
\begin{equation}	
F =\frac{Gm}{\Delta^2} \sqrt{1+2u^2\cos\varphi+u^4} \label{eqn:perfturbF}\quad ,
\end{equation}	
and 
\begin{equation}
\begin{array}{llll}
F_1=\frac{GM}{\Delta^2 (1+2u\cos\varphi)} \Bigg[ & 1+ \bigg(1+2u\cos\varphi + u^2 \bigg)^2 \\
	& -2(1+u\cos\varphi)\sqrt{1+2u\cos\varphi+u^2} \Bigg]^{\nicefrac{1}{2}} \quad . \label{eqn:perfturbF1}
\end{array} 
\end{equation}
For step-by-step calculations to arrive to Equations (\ref{eqn:perfturbF}) and (\ref{eqn:perfturbF1}), please refer to \citep{1964SvA.....7..618C}\footnote{Note that there is a typo in the expression of $F_1$ given in \citep{1964SvA.....7..618C}, with an extra factor $u$ on the right side of this equation (number 18 in the paper). However, this typo is not carried out in any calculations, neither before nor after that equation.}.

In what follows, we keep in mind the geometry of the problem illustrated in Fig. \ref{fig:3bp-rep}, {i.e.} $r_1>>\Delta$. Thus from Equation (\ref{eqn:u}), we have $u<<1$. The expressions of the four aforementioned perturbations simplify as follows: 
\begin{equation}	
R = \frac{GM}{r^2}  \label{eqn:PerturbationR}\quad ,
\end{equation}				
\begin{equation}	
       F= R_1 = \frac{Gm}{\Delta^2} \label{eqn:PerturbationR1ouF} \quad ,
\end{equation}	
and 
\begin{equation}
       F_1 = \frac{GM\Delta}{r_1^3} \sqrt{1+3\cos^2\varphi} \quad . \label{eqn:PerturbationF1}
\end{equation}		

\textbf{Note:} The expression of $F_1$ given by Equation (\ref{eqn:PerturbationF1}) is truncated to the 0$^{th}$ order in $u$ (cf. Equation \ref{eqn:perfturbF1}), this is more than sufficient for our analysis.

In all that follows, we will exclusively be using the expressions given by equations (\ref{eqn:PerturbationR}) to (\ref{eqn:PerturbationF1}).
\vspace{-0.3cm}

\section{Implications of the accelerations on the physical and geometrical boundaries}
\vspace{-0.1cm}
\subsection{Boundary conditions: the sphere of activity} 
\label{sect:BoundaryConditions}
\vspace{-0.2cm}
In his 1964 paper, G.A. Chebotarev defines the sphere of activity of a planet and derives its boundaries. He defines the \textit{spheres of activity} as follows: 
\begin{quotation} 
\textit{"Sphere of activity" refers to that region of space in which it is feasible to assume a planet as the central body and the Sun as the perturbing body when computing perturbations.}
\end{quotation}

In this paper, we revisit the different spheres of (gravitational) activity, to which we give a geometrical meaning. 

From the accelerations given by Equations (\ref{eqn:PerturbationR}) to (\ref{eqn:PerturbationF1}), equating the ratio of Solar accelerations (central/perturbing) to that of the planetary accelerations (central/perturbing) gives the natural bounding surface of the sphere of activity:
\begin{equation}
\frac{F}{R}  =  \frac{F_1}{R_1} \label{eqn:ActSphere} \quad ,
\end{equation}
within the sphere of activity
\begin{equation}
\frac{F}{R}  >  \frac{F_1}{R_1}  \label{ineqn:ActSphere} \quad .
\end{equation}

Substituting Equations (\ref{eqn:PerturbationR}) to (\ref{eqn:PerturbationF1}) into (\ref{eqn:ActSphere}), we write that the radius $\Delta$ of the surface of activity is equal to $\Delta_{1}$: 
\begin{equation}
\Delta_{1} = \Bigg[ \frac{\overline{m}}{\sqrt{1+3\cos^2\varphi}}\Bigg]^{2/5} \cdot r_1 	\label{eqn:RadiusActSphere} \quad ,
\end{equation}
where $\overline{m}=\frac{m}{M}$ denotes the normalised mass of the planet.

Equation (\ref{eqn:RadiusActSphere}) delimits in polar coordinates the hyper-surface of rotation which defines the borders of the so-called "sphere of activity" at a given heliocentric distance of the planet.
\vspace{-0.1cm}

\subsection{Analysis of the planetary accelerations and the ratio ${\nicefrac{F_1}{R_1}}$}
\label{sect:ratioAnalysis}
\vspace{-0.1cm}
We take a closer look at the individual planetary accelerations involved, as well as their ratio $\frac{F_1}{R_1}$ (central/perturbing) by substituting Equations (\ref{eqn:PerturbationR1ouF}) and (\ref{eqn:PerturbationF1}) into $\nicefrac{F_1}{R_1}$, we write
\begin{equation}
\frac{F_1}{R_1} = \frac{\Delta^3}{\overline{m}\,r_1^3} \sqrt{1+3\cos^2\varphi} \quad . \label{eqn:ratioF1R1general}
\end{equation}

In particular, for a given system (with given $\overline{m}$), we consider the following:
\begin{enumerate}
\item the maximum of the ratio $\frac{F_1}{R_1}$:\\
\vspace{-0.2cm}

 From Equation (\ref{eqn:ratioF1R1general}), we maximise the ratio for-all $\Delta$ and for-all $r_1$ that is satisfied when $\cos^2\varphi=1$. From Fig. \ref{fig:3bp-rep} this would mean an alignment from left to right Sun-Planet-Satellite or Sun-Satellite-Planet, i.e. full moon or new moon, respectively.  Thus, the maximum of the ratio is 
\begin{equation}
\frac{F_1}{R_1} = 2\,\sqrt[5\,]{\,\overline{m}\,} \label{eqn:ratioF1R1} \quad .
\end{equation}
This relation gives us an estimate of the maximum ratio $\frac{F_1}{R_1}$ on the bounding surface of the sphere of activity. As can be seen from Equation (\ref{eqn:ratioF1R1}), this ratio is solely dependent on the normalised mass and is completely independent of the heliocentric distances and the geometry of the orbit.\\

\item the region where $F_1=R_1$:\\
\vspace{-0.2cm}

In Equation (\ref{eqn:ratioF1R1general}), the ratio ${\nicefrac{F_1}{R_1}}=1$ gives:
 \begin{equation}
\Delta_2 =  r_1 \Bigg[\frac{\overline{m}}{\sqrt{1+3\cos^2\varphi}}\Bigg]^{\nicefrac{1}{3}} \label{eqn:DeltaF1eqR1} \quad .
\end{equation}
Equation (\ref{eqn:DeltaF1eqR1}) delimits in polar coordinates the hyper-surface for which $F_1=R_1$. Just like in the previous section with the sphere of activity, this hyper-surface is not quite a sphere.

At a given planetary heliocentric distance, $r_1$, we write the 
\begin{itemize}
\item[ ] maximum radius
\begin{equation}
\quad \Delta_{2_\text{\,sup}} = r_1\, \sqrt[3\,]{{\overline{m}}}\quad$ $\Big(\varphi = (k+\frac{1}{2})\,\pi \, , \,k\in\mathbb{Z}\Big) \label{eqn:Dealta2Sup} \quad ,
\end{equation}

\item[] and minimum radius
\begin{equation}
\quad\Delta_{2_\text{\,inf}} = r_1\, \sqrt[3\,]{\frac{\overline{m}}{2}}\quad$  $\Big(\varphi = k\,\pi \, , k\in\mathbb{Z}\Big)\quad.  \label{eqn:Dealta2Inf}
\end{equation}
\end{itemize}

The ratio $\nicefrac{\Delta_{2_\text{\,sup}}}{\Delta_{2_\text{inf}}}$ of these extremal values is at all times and for all systems constant, and is equal to $\sqrt[3\,]{2}$.
\end{enumerate}

\section{Gravitational Spheres}

\subsection{Laplace Sphere}
\label{subsect:LaplaceSphere} 
As aforementioned, \textsc{Laplace}, P. S. was the first to tackle the problem of spheres of influence of planets. He was the first to derive the radius of the sphere of activity of a planet (see pages 217 - 219 of \citep{Laplace1878}).

 Following \citep{Laplace1878} and \citep{1964SvA.....7..618C}, we elaborate on the properties of the sphere of activity. We also shed some light on the confusion that might exist in the literature between the different definitions of spheres of activity.

Equation (\ref{eqn:RadiusActSphere}) gives the expression of the hyper-surface of rotation which delimits the "volume of gravitational activity of a planet". At each point along the planet's orbit, the so-called sphere is centred at the point-mass planet.

Although, this hyper-surface is not a sphere \textit{per se}, one can define a homocentric sphere along each point on the orbit (i.e. fixed value of heliocentric distance $r_1$). We write the associated:
\vspace{-.20cm}
\begin{itemize}
\item[] maximum radius
\vspace{-.1cm}
\begin{equation}
\Delta_{1_\text{\,sup}} = r_1\, \sqrt[5\,]{{\overline{m}}^{2}}\quad$ $\Big(\varphi = (k+\frac{1}{2})\,\pi \, , k\in\mathbb{Z}\Big) \label{eqn:Dealta1Sup} \quad ,
\end{equation}

\item[] and minimum radius
\vspace{-.1cm}
\begin{equation}
 \Delta_{1_\text{\,inf}} = r_1\, \sqrt[5\,]{\frac{\overline{m}^2}{2}}\quad$  $\Big(\varphi = k\,\pi \, , k\in\mathbb{Z}\Big) \quad.  \label{eqn:Dealta1Inf}
\end{equation}
\end{itemize}
\vspace{-.1cm}

The ratio $\,\nicefrac{\Delta_{1_{_\text{sup}}}}{\Delta_{1_{_\text{inf}}}}$ of these extremal values is $ \sqrt[5\,]{2}$.\\

For a given heliocentric distance of the planet $r_1$, what is usually referred to as the \underline{\textit{radius of sphere of activity}} of the planet is the maximum of these two radii (given by Equations (\ref{eqn:Dealta1Sup}) and (\ref{eqn:Dealta1Inf})). That is $\Delta_{1_{sup}}$ which is given by Equation (\ref{eqn:Dealta1Sup}) (i.e. $\,\Delta_{1}$ when $\varphi= (k+\frac{1}{2})\,\pi \,\, ,\, \,\,k\in\mathbb{Z}\, \implies\,\bf{r_1} \perp \bf{r'}$).\\
\vspace{-.3cm}

For the rest of this paper and in our future works, we will note $R_{_\text{Laplace}}$ the radius of the \textit{gravitational sphere of activity}\footnote{written $\Delta_1$ in \citep{1964SvA.....7..618C}}:
\vspace{-.15cm}
 \begin{equation}
R_{_\text{Laplace}}= \Delta_{1_{_\text{sup}}} (r_1)= r_1\, \overline{m}^{\nicefrac{2}{5}} \label{eqn:RayonSphereLaplace} \quad . %
\end{equation}
\vspace{-.1cm}
This is the expression given in \citep{Laplace1878} for the sphere of activity.

\subsubsection{Discussion}
  As the planet revolves around its host star, its heliocentric distance $r_1$ varies between two extremal values at the planet's perihelion $r_{1_{_\text{min}}}=a\,(1-e)$ and planet's aphelion $r_{1_{_\text{max}}}=a\,(1+e)$. Where $a$ and $e$ are the semi-major axis and the eccentricity of the planet, respectively.

The two extremal values of the radius $R_{\textit{Laplace}}$ are: \\
 at the planet's perihelion 
 \begin{equation}
 R_{{\text{Laplace}_\text{\,min}}} = R_{\text{Laplace}}\Big|_\text{perihelion}= a(1-e)\,\overline{m}^{\,\nicefrac{2}{5}}  \label{eqn:RayonSphereMinPeri} \quad ,
 \end{equation}
 and at the planet's aphelion 
 \begin{equation}
 R_{{\text{Laplace}_{\,max}}} = R_{\text{Laplace}}\Big|_\text{aphelion} = a(1+e)\,\overline{m}^{\,\nicefrac{2}{5}}  \quad . \label{eqn:RayonSphereMaxAph}
 \end{equation}

\begin{figure*}
  \centering
 \includegraphics[width=15.7cm]{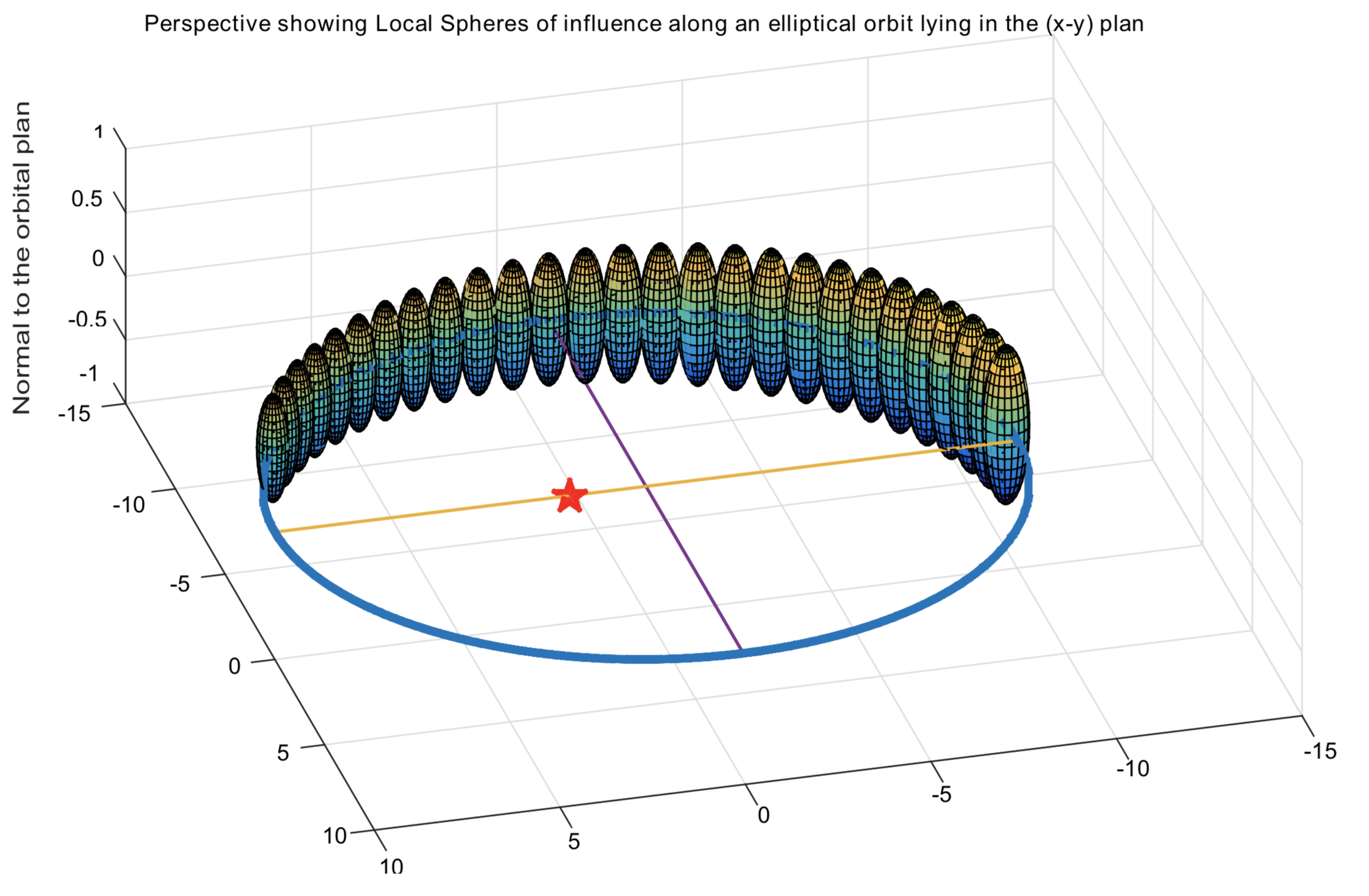}\\
   \caption{\label{fig:EllipticalOrbit}Global behaviour of local activity spheres along an elliptical orbit ($e= 0.2$). The Sun is represented by a red star at a focal point of the ellipse. The blue bold line represents the planet's orbit, whereas each sphere (of activity) is centred at a point along the orbit.    }
\end{figure*}

We represent in Fig. \ref{fig:EllipticalOrbit} successive spheres of activity along an elliptical orbit for Mercury ($e=0.2$ and $\overline{m}\sim 1.65\times10^{-7}$). The distances $r_1$ and the radii $\Delta_1$ are not to scale for the sake of clarity of the figure.

As seen in Fig. \ref{fig:EllipticalOrbit}, and expected from Equation (\ref{eqn:RayonSphereLaplace}), the local spheres with the minimal radii are at perihelia (Equation \ref{eqn:RayonSphereMinPeri}); whereas those with maximal radii are at aphelia (Equation \ref{eqn:RayonSphereMaxAph}).

\subsubsection{Evaluating the ratio $\nicefrac{F_1}{R_1}$ at the Laplace Radius}
\label{sect:F1R1atLaplace}
Using Equation (\ref{eqn:ratioF1R1general}), we compute $\nicefrac{F_1}{R_1}$ at the Laplace Radius (given by Equation \ref{eqn:RayonSphereLaplace}). We write

\begin{equation}
\frac{F_1}{R_1}{\Bigg|_{R_{_\text{Laplace}}}}= { \frac{\Delta^3}{\overline{m}\,r_1^3} \sqrt{1+3\cos^2\varphi} }\,\,\Bigg|_{R_{_\text{Laplace}}} \quad .
\end{equation}

We are interested in the maximum of this ratio, i.e. with $\cos^2 \varphi=1$, which gives
\begin{equation}
\frac{F_1}{R_1}{\Bigg|_{R_{_\text{Laplace}}}}=2\,(\,\overline{m}\,)^{\nicefrac{-3}{5}} \quad .
\end{equation}

This relation gives us an estimate of the maximum ratio $\frac{F_1}{R_1}$ on the bounding surface of the Laplace sphere. This ratio is solely dependent on the ratio of the masses and is completely independent of the heliocentric distances and the geometry of the orbit.
\vspace{-.1cm}

\subsubsection{The case of circular planetary orbits}
\label{subsucsec:circular}
In case of a circular or nearly circular orbit for the planet ($e<<1$), Equation (\ref{eqn:RayonSphereLaplace}) writes 
\begin{equation}
R_{_{_\text{Laplace}}}= a\,\overline{m}^{\,\nicefrac{2}{5}}   \label{eqn:RayonSphereCircularOrbit} \quad .
\end{equation}

\begin{figure*}
  \centering
 \includegraphics[width=15.7cm]{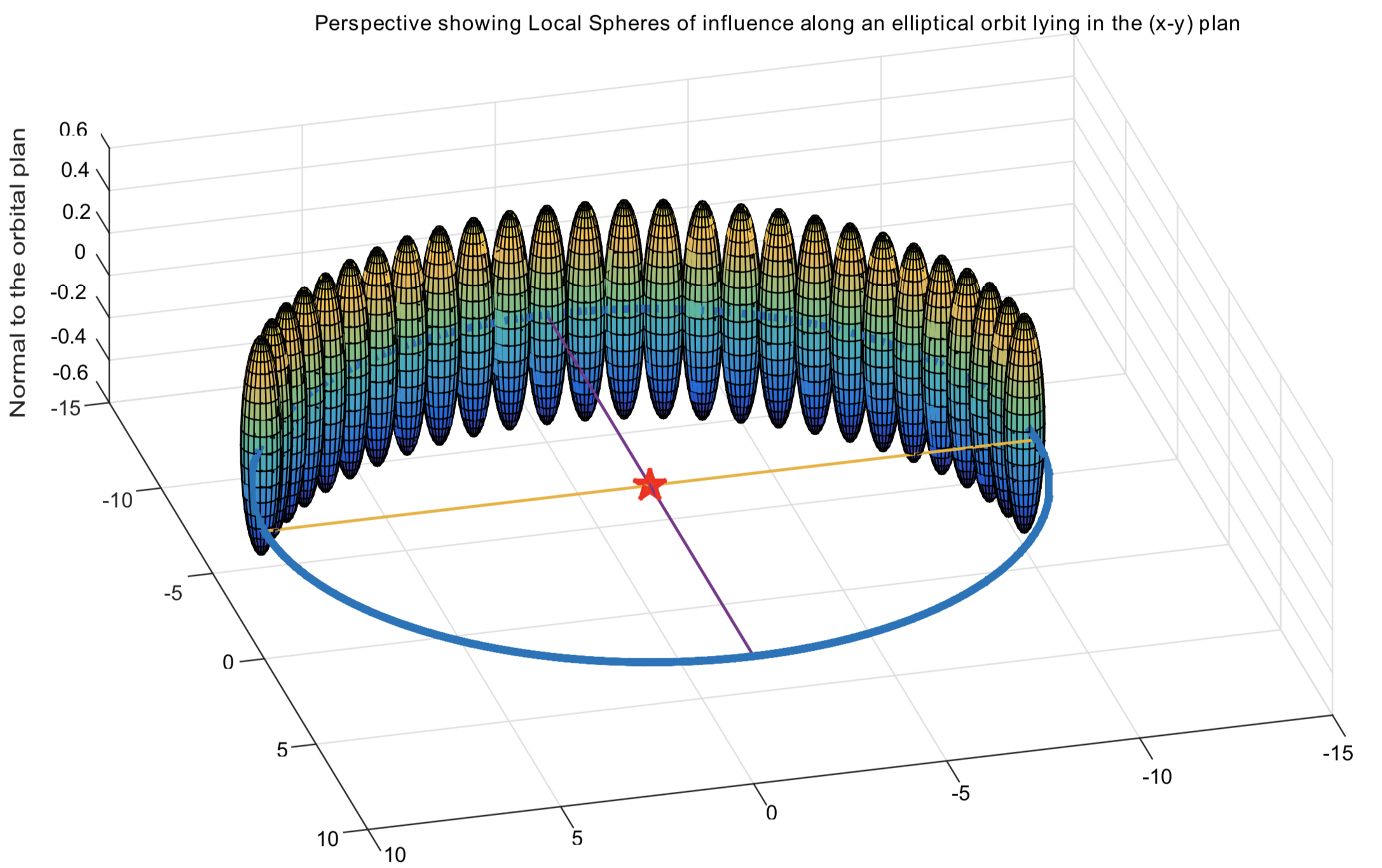}\\
   \caption{\label{fig:torus}Global behaviour of local activity spheres for a planet along a circular orbit. The Sun is represented by the star. The blue bold line represents the planet's orbit, whereas each sphere (of activity) is centred at a point along the orbit. The global behaviour of these spheres of equal radii draws a torus. 
   }
   \textbf{Note:} In Fig. \ref{fig:EllipticalOrbit} and Fig. \ref{fig:torus}, each sphere of influence is coloured solely for the sake of legibility. Each colour is associated with a certain value of geo-centric distance $\Delta$ and a specific value of $\sin\varphi$.
\end{figure*}

Fig. \ref{fig:torus} shows the local spheres of activity along a circular orbit (e.g. a fictitious Mercury on a circular orbit). In this case, as the planet revolves around its host star on a circular orbit, the successive Laplace spheres draw a torus along the orbit, what we call the Laplace Torus (cf. Fig. \ref{fig:torus}). 

\subsection{Chebotarev Sphere}
\label{sect:CheboSphere}
As per \citep{1964SvA.....7..618C}, the hyper-surface bounding the gravitational sphere of influence is given by $R_1=R$, {i.e.}
\begin{equation}
 \frac{M}{r^2} = \frac{m}{\Delta^2}  \quad .
 \end{equation}

Within what \cite{1964SvA.....7..618C} calls the gravitational sphere 
\begin{equation}
R_1>R \label{eqn:CheboSphereIneq} \quad .
\end{equation}
We will call this region the Chebotarev Sphere.

Substituting Equations (\ref{eqn:PerturbationR}) and (\ref{eqn:PerturbationR1ouF}) into in-equation (\ref{eqn:CheboSphereIneq}), the Chebotarev Sphere\footnote{denoted $\Delta_2$ in \citep{1964SvA.....7..618C},} is:
\begin{equation}
\Delta_3 < r_1 {\,\overline{m}\,}^{\nicefrac{1}{2}}  \label{eqn:CheboSphere} \quad .
\end{equation}
We write the Chebotarev radius
\begin{equation}
R_{_\text{Ch}} = r_1 {\,\overline{m}\,}^{\nicefrac{1}{2}}  \label{eqn:CheboRadius} \quad .
\end{equation}


\subsubsection*{- Evaluating the ratio ${\nicefrac{F_1}{R_1}}$ at the Chebotarev Radius}
\label{sect:F1R1atChebotarev}
Using Equation (\ref{eqn:ratioF1R1general}), we compute $\nicefrac{F_1}{R_1}$ at the Chebotarev Radius (given by Equation (\ref{eqn:CheboRadius})). We write
\begin{equation}
\frac{F_1}{R_1}{\Bigg|_{R_{_\text{Ch}}}}= { \frac{\Delta^3}{\overline{m}\,r_1^3} \sqrt{1+3\cos^2\varphi} }\,\,\Bigg|_{R_{_\text{Ch}}} \quad .
\end{equation}

We are interested in the maximum of this ratio, {i.e.} with $\cos^2 \varphi=1$, which gives
\begin{equation}
\frac{F_1}{R_1}{\Bigg|_{R_{_\text{Ch}}}}=2\,\sqrt{\,\overline{m}\,} \quad .
\end{equation}

This relation gives us an estimate of the maximum ratio $\frac{F_1}{R_1}$ on the bounding surface of the Chebotarev sphere. This ratio is solely dependent on the ratio of the masses and is completely independent of the heliocentric distances and the geometry of the orbit.

\vspace{-0.2cm}
\subsection{Hill Sphere}
\label{sect:HillSphere}
\vspace{-0.1cm}
The Hill Sphere is the most commonly used gravitational sphere in celestial mechanics. It is defined as the region around a planet within which the satellite's motion around the planet is stable and where the planet's gravity dominates over that of the star (of mass $M$). It extends between the Lagrangian libration points $L_1$ and $L_2$.

The problem has been extensively investigated in the literature (\citealt{1988ahl..book.....R} and references therein). In our notations, we write (to the first order in $\bar{m}$)
\begin{equation}
R_{_\text{Hill}} = a \, \sqrt[3\,]{\frac{m}{3\,M}} =  a \, \sqrt[3\,]{\frac{\bar{m}}{3}} \label{eqn:HillRadius}  \quad .
\end{equation}
 
Though the definition of the \textit{Hill Sphere} might resemble, at first glance, that of the \textit{Chebotarev Sphere}, it is much more restrictive from a dynamical point of view as the stability of the orbits is required.

\vspace{-0.2cm}
\subsubsection{Evaluating the ratio ${\nicefrac{F_1}{R_1}}$ at the Hill Radius}
\label{sect:F1R1atHill}
\vspace{-0.1cm}
Using Equation (\ref{eqn:ratioF1R1general}), we compute $\nicefrac{F_1}{R_1}$ at the Hill Radius (given by Equation \ref{eqn:HillRadius}). We write
\begin{equation}
\frac{F_1}{R_1}{\Bigg|_{R_{_\text{Hill}}}}=  { \frac{\Delta^3}{\overline{m}\,r_1^3} \sqrt{1+3\cos^2\varphi} }\,\,\Bigg|_{R_{_\text{Hill}}} \quad ,
\end{equation}
with $\cos^2 \varphi=1$, which gives
\begin{equation}
\frac{F_1}{R_1}{\Bigg|_{R_{_\text{Hill}}}}\le \frac{2}{3}\quad .
\end{equation}

We are interested in the maximum of the ratio $\nicefrac{F_1}{R_1}$ at the surface of the Hill Sphere, which we evaluate when $\cos^2 \varphi=1$:
\begin{equation}
\frac{F_1}{R_1}{\Bigg|_{R_{_\text{Hill}}}}= \frac{2}{3}  \quad .
\end{equation}

Unlike the ratio $\frac{F_1}{R_1}$ for the sphere of activity (or the Laplace sphere which is a particular case of the sphere of activity, or the Chebotarev) which depend on the reduced mass ($\overline{m}$), this ratio evaluated at Hill radius is at all times constant and independent of the masses involved.

\vspace{-0.2cm}
\subsubsection{Discussions}
\label{sect:discussions}
\vspace{-0.1cm}
\begin{enumerate}
\item \textit{Comparison of the Hill sphere to the sphere of radius $\Delta_2$}
 \end{enumerate}
 \vspace{-0.2cm}

As seen in the point (ii) of Section \ref{sect:ratioAnalysis}, $F_1=R_1$ defines a hyper-surface of radius $\Delta_2$ given by Equation (\ref{eqn:DeltaF1eqR1}). For a given planetary heliocentric distance, the $\Delta_2$ max is given by Equation (\ref{eqn:Dealta2Sup}). At~$r_1~=~a$, we write:
$$\Delta_{2}(r_1=a) = a \, \sqrt[3\,]{\bar{m}} \quad . $$ 
We write the ratio of $\Delta_{_2}(a)$ and $R_{_\text{Hill}}$:
\begin{equation}
\frac{\Delta_{_2}}{R_{_\text{Hill}}} = \frac{a \, \sqrt[3\,]{\bar{m}}} { a \, \sqrt[3\,]{\frac{\bar{m}}{3}} } =  \sqrt[3\,]{3} \approx 1.44 \quad .
\end{equation}
In other words, $\Delta_{_2}$, the radius of the gravitational sphere (bounding $\nicefrac{F_1}{R_1}=1$) is about 1.44 times greater than the Hill radius $R_{_\text{Hill}}$. \\
At any given planetary heliocentric distance, $r_1$, we write:
\begin{equation}
\Delta_{_2} \approx 1.44\,\,{R_{_\text{Hill}}}  \quad .
\end{equation}
\vspace{0.2cm}

\begin{enumerate}
\setcounter{enumi}{1}
\item \textit{Laplace Sphere vs. Hill Sphere}
 \end{enumerate}
 \vspace{-0.1cm}
 
The ratio of the Laplace Sphere (Equation \ref{eqn:RayonSphereLaplace}) to that of the Hill Sphere (Equation \ref{eqn:HillRadius}) writes
\begin{equation}
\frac{R_{_\text{Hill}}}{R_{_\text{Laplace}}}=\frac{ a \, \sqrt[3\,]{\frac{\bar{m}}{3}}}{r_1\,\overline{m}^{\nicefrac{2}{5}}}=\frac{ a \, \sqrt[3\,]{\frac{\bar{m}}{3}}}{a\,\overline{m}^{\nicefrac{2}{5}}} \quad .
\end{equation}

Typically, in the case of bound orbits which are of interest for us (e.g. planet of mass $m$ revolving around a star of mass $M$, or a star of mass $m$ revolving around a galactic centre of mass $M$), we have~$e<<1$. Therefore it is safe to replace $r_1$ by $a$, and $\bar{m}<<1$. 
\begin{equation}
\implies R_{_\text{Hill}}>R_\text{Laplace}\quad .
\end{equation}
\vspace{0.2cm}

\begin{enumerate}
\setcounter{enumi}{2}
\item  \textit{Discussion: Chebotarev Sphere vs. Laplace Sphere}
\end{enumerate}
 \vspace{-0.2cm}
 
Finally, we write the ratio of the Chebotarev Sphere (Equation \ref{eqn:CheboRadius}) to that of the Laplace Sphere (Equation \ref{eqn:RayonSphereLaplace})
 \begin{equation}
\frac{R_{_\text{Ch}}}{R_\text{Laplace}}=\frac{ \cancel{r_1} \,\sqrt{\bar{m}}}{\cancel{r_1}\,\overline{m}^{\nicefrac{2}{5}}}= \sqrt[10\,]{\overline{m}}\quad .
\end{equation}
This ratio depends solely on the reduced mass $\overline{m}$. In the systems that are of interest for us, $\overline{m}<<1$, this implies that 
  \begin{equation}
R_{_\text{Ch}}<R_{_\text{Laplace}} \quad .
\end{equation}
\vspace{0.2cm}

\begin{enumerate}
\setcounter{enumi}{3}
\item To summarise, at a given $r_1$:
\end{enumerate}
 \vspace{-0.3cm}
 \begin{equation}
R_{_\text{Ch}}<R_{_\text{Laplace}} < R_{_\text{Hill}}< \Delta_{_2} \quad .
\end{equation}

The general statement given by the previous equation is in agreement with the numerical results given in \cite{1964SvA.....7..618C} and in the rest of this paper.

\begin{table*}
\begin{center}
\begin{tabular}{SlScScScScScScSc}
\hline %
\multirow{2}{*}{at a given $r_1$ (in our notations)}	&  \multirow{2}{*}{$R_{_\text{Ch}}$} &  \multirow{2}{*}{<}  &  \multirow{2}{*}{$R_\textit{Laplace} $}  &  \multirow{2}{*}{< }&  
\multirow{2}{*}{$ R_{_\text{Hill}}$} &  \multirow{2}{*}{< }&  \multirow{2}{*}{$\Delta_{_2}$} \\		
	&   &   &    & & & & 	\\			
	\hline	
 \multirow{2}{*}{Physical meaning}	&   \multirow{2}{*}{($R_1=R$)} & &   \multirow{2}{*}{$\bigg(\frac{F}{R}>\frac{F_1}{R_1}\bigg)$}  & & \multirow{2}{*}{ stability required} & &  \multirow{2}{*}{ ($R_1=F_1$)}	\\
	&   &   &    & & & & 	\\	%
\multirow{1}{*}{In \cite{1964SvA.....7..618C}'s } &   \multirow{2}{*}{$\Delta_{2}$} &   & \multirow{2}{*}{$\Delta_{1}$}   & & \multirow{2}{*}{$\Delta_{3}$}& & \multirow{2}{*}{$\Delta_{4}$} \\	
notations		&   &   &    & & & & 	\\
\hline %
\end{tabular}
\caption{\label{tab:summary2}Ordering of the different radii studied in this paper, and the associated physical meaning.}
\end{center}
\end{table*}

 \vspace{-0.1cm}
\begin{table*}
\begin{center}
\begin{tabular}{SlScSc|ScSc}
\hline
   \multicolumn{3}{c}{This paper} & \multicolumn{2}{l}{\cite{1964SvA.....7..618C}} \\
 name & Section	& expression  & symbol   & name \\		
\hline %
 \multirow{3}{*}{Surface of activity}	&  \multirow{3}{*}{ \ref{sect:BoundaryConditions} } &   \multirow{3}{*}{$\Delta_{1} = \Bigg[ \frac{\overline{m}}{\sqrt{1+3\cos^2\varphi}}\Bigg]^{2/5} \cdot r_1 $}  &  \multirow{3}{*}{$\Delta_{1}$} &   \multirow{3}{*}{Surface of activity} \\
	& & & &  \\		
	& & & &  \\			
   \multicolumn{5}{c}{\multirow{2}{*}{Within the sphere of activity: $\frac{F}{R}> \frac{F_1}{R_1}$} }\\
      \multicolumn{5}{c}{	} \\		   %
 \multirow{3}{*}{Laplace sphere} &  \multirow{3}{*}{\ref{subsect:LaplaceSphere} }  &  \multirow{3}{*}{ $R_\textit{Laplace}= r_1\, \overline{m}^{\nicefrac{2}{5}}$}   &   \multirow{3}{*}{$\Delta_{1}$}   & \multirow{2}{*}{The maximum radius of surface of rotation.}	\\		
	&  &   &    &	\\	
	&  &   &    & (associated to the sphere of activity) \\		
      \multicolumn{5}{c}{	} \\	%
 \multirow{2}{*}{Chebotarev Sphere}	&  \multirow{2}{*}{\ref{sect:CheboSphere}}	&  \multirow{2}{*}{$\Delta_3 < r_1 {\,\overline{m}\,}^{\nicefrac{1}{2}} $} &  \multirow{2}{*}{-} &  \multirow{2}{*}{Gravitational sphere } \\
	&  &   &    &	\\	
   \multicolumn{5}{c}{\multirow{2}{*}{Within the Chebotarev sphere: $R_1> R$} }\\
      \multicolumn{5}{c}{	} \\		%
 \multirow{2}{*}{Chebotarev Radius}	&  \multirow{2}{*}{\ref{sect:CheboSphere}}	&  \multirow{2}{*}{$R_{_\text{Ch}} < r_1 {\,\overline{m}\,}^{\nicefrac{1}{2}} $}  &  \multirow{2}{*}{$\Delta_2$} &  \multirow{2}{*}{Radius of the Gravitational sphere} \\
	&  &   &    &	\\	
      \multicolumn{5}{c}{	} \\		%
\multirow{3}{*}{Hill Sphere}	& \multirow{3}{*}{\ref{sect:HillSphere}} & \multirow{3}{*}{$R_{_\text{Hill}} = a \, \sqrt[3\,]{\frac{m}{3\,M}} =  a \, \sqrt[3\,]{\frac{\bar{m}}{3}} $}  &  \multirow{3}{*}{$\Delta_3$}   &  \multirow{2}{*}{Hill Sphere}	\\
	&  &   &    & \\	
	&  &   &    & (The stability of the orbits is required here.)		\\	      		
\hline%
   \multicolumn{5}{c}{	} \\
      \multicolumn{5}{c}{	} \\	 
\hline%
   \multicolumn{5}{c}{\textbf{Analysing the accelerations: main results}} \\	 
\hline%
   \multicolumn{3}{l}{ \multirow{5}{*}{The ratio of planetary acceleration (central/perturbing)}} &  
    \multicolumn{2}{l}{ \multirow{5}{*}{$\frac{F_1}{R_1} = \frac{\Delta^3}{\overline{m}\,r_1^3} \sqrt{1+3\cos^2\varphi} \quad $ (cf. Section \ref{sect:ratioAnalysis})}} \\	
   \multicolumn{5}{c}{	} \\	         
   \multicolumn{5}{c}{	} \\	   
   \multicolumn{3}{l}{ \multirow{3}{*}{On the bounding surface of the sphere of activity the maximum ratio}} &
   \multicolumn{2}{l}{\multirow{3}{*}{$\frac{F_1}{R_1}\Big|_{\cos^2\varphi=1}  = 2\,\sqrt[5\,]{\,\overline{m}\,}  \qquad $ (cf. Section \ref{sect:ratioAnalysis}) }} \\	
   \multicolumn{5}{c}{	} \\	        
   \multicolumn{5}{c}{	} \\	   
   \multicolumn{3}{l}{ \multirow{3}{*}{Surface on which $R_1=F_1 $ (cf. Section \ref{sect:ratioAnalysis})}} &
   \multicolumn{2}{l}{ \multirow{3}{*}{$ \Delta_2 = r_1 \Bigg[ \frac{\overline{m}}{\sqrt{1+3\cos^2\varphi}}\Bigg]^{1/3} $ $\quad$ (denoted $\Delta_4$ in \cite{1964SvA.....7..618C}) } }\\	
   \multicolumn{5}{c}{	} \\	   
   \multicolumn{5}{c}{	} \\	        
   \multicolumn{3}{l}{ \multirow{3}{*}{  At the Laplace radius: }} &
    \multicolumn{2}{l}{ \multirow{3}{*}{   $\textit{max}\,\frac{F_1}{R_1}{\Bigg|_{R_{_\textit{Laplace}}}} =  2\,(\,\overline{m}\,)^{^{\nicefrac{-3}{5}}}\qquad$ (cf. Section \ref{subsect:LaplaceSphere}) } } \\ 
   \multicolumn{5}{c}{	} \\	   
   \multicolumn{5}{c}{	} \\	        
   \multicolumn{3}{l}{ \multirow{3}{*}{  At the Chebotarev radius: }} &
   \multicolumn{2}{l}{ \multirow{3}{*}{     $\textit{max}\,\frac{F_1}{R_1}{\Bigg|_{R_{_\text{Ch}}}} = 2\,\sqrt{\,\overline{m}\,} \qquad$ (cf. Section \ref{sect:CheboSphere}) }} \\	
   \multicolumn{5}{c}{	} \\	   
   \multicolumn{5}{c}{	} \\	        
   \multicolumn{3}{l}{ \multirow{3}{*}{   At the Hill radius:}} &
    \multicolumn{2}{l}{ \multirow{3}{*}{  $\textit{max}\,\frac{F_1}{R_1}{\Bigg|_{R_{_\text{Hill}}}} \le \frac{2}{3}  \qquad $ (cf. Section \ref{sect:F1R1atHill}) }} \\	   
   \multicolumn{5}{c}{	} \\	        
   \multicolumn{5}{c}{	} \\	   
\hline %
\end{tabular}
\caption{\label{tab:summary}Table summarising the different concepts of gravitational spheres of influence studied in this paper (surface of activity, sphere of activity, Laplace sphere, Chebotarev sphere, and Hill sphere).}
\end{center}
\end{table*}

\vspace{-0.2cm}
\section{The Roche Limit}
\label{sect:ROcheLimit}
The Roche limit is defined as the distance from the centre of a planet within which any large satellite would be torn apart because of tidal forces.  It is named after \'Edouard Albert \textsc{Roche} (1820 - 1883) who introduced the concept in 1848.

\textsc{Roche}, E. A. investigated the problem of the gravitational processes at the origin of planetary rings around the giant planets. The effect of tidal forces are crucial here, as they can lead to the destruction of a satellite when too close to the planet.
 
 Let us assume the following system of a planet + satellite at distance $r'$ from one another. We denote by $M_\text{planet}$ and $M_\text{sat}$ the masses of the planet and the satellite, respectively. $R_\text{planet}$ and $R_\text{sat}$ are their respective radii. In what follows, we work in a planeto-centric, non-Galilean reference frame.
 
 We define $\mathbf{u}$ as a unit vector along the direction centre of mass of the planet and that of the satellite, and pointing towards the satellite. Let us consider a point \textit{P} of mass $m$ in the surface of the satellite. The forces acting on the point-mass $P$ are the following:
 \begin{itemize}
\item the gravitational forces 
  \begin{equation}
 \mathbf{F}_\text{planet\,/\,P} =\frac{-GmM_\text{planet}}{(r'+R_\text{sat})^2}\,\mathbf{u} \quad ; \quad  \mathbf{F}_\text{sat\,/\,P} =\frac{-GmM_\text{sat}}{R^2_\text{sat}}\mathbf{u} \quad ,
 \end{equation}

\end{itemize}
\begin{itemize}
 \item the Coriolis forces: they are nil at equilibrium.\\
 \item the inertial force: written as follows: 
   \begin{equation}
 \mathbf{F} =m\,(r'+R_\text{sat})\,\omega^2\,\mathbf{u} \qquad \textrm{where} \qquad \omega^2=\frac{GM_\text{planet}}{r^3} \quad .
 \end{equation}
 \end{itemize}
 Applying the parallel axis theorem gives us $\mathbf{R}$ (the reaction of the satellite)
    \begin{equation}
 \mathbf{R} =G\,m\,R_\text{sat}\, \bigg(\frac{M_\text{sat}}{R^3_\text{sat}} + \frac{M_\text{planet}}{r^3}\bigg)\,\mathbf{u} \label{eqn:reaction} \quad .
 \end{equation}
 
The condition for the satellite not to be torn apart and shattered, and for the point-mass \textit{P} to belong to the satellite, requires $R>0$.
Thus from Equation (\ref{eqn:reaction}), we write that 
    \begin{equation}
r' > R_\text{planet} \sqrt[3\,]{3\frac{\mu_\text{planet}}{\mu_\text{sat}}}  \quad ,
 \end{equation}
 where $\mu_\text{planet}$ and $\mu_\text{sat}$ are the bulk densities of the planet and satellite, respectively.
 
 To be more complete in our analysis, we would have to consider the tensile strengths. In that case, the previous equation gives
 \vspace{-0.2cm}
     \begin{equation}
r' > R_\text{planet} 2.46 \sqrt[3\,]{\frac{\mu_\text{planet}}{\mu_\text{sat}}} \label{eqn:Roche} \quad .
 \end{equation}
 From Equation (\ref{eqn:Roche}), we can say that in the case where the planet and the satellite have similar densities, the Roche limit lies at about 2.46 planetary radii.
 
 When the orbiting body crosses the Roche limit (critical distance given by Equation (\ref{eqn:Roche}), tidal forces across the body are much stronger than the cohesive forces holding the body together, resulting in the shattering of the satellite. \\
Some exceptions do exist within our solar system. We can cite Jupiter's moon Metis (semi-major axis $128\times10^3\,$km, Jupiter's radius $\approx71\times10^3$\,km, that is a ratio of about 1.8), and Saturn's moon Pan (semi-major axis of $134\times10^3\,$km. Saturn's radius $\approx7160\times10^3$\,km, that is a ratio of about 2.2). These moons hold together because of their tensile strength. We present more exceptions in the Giant planets system in the Appendix.
\vspace{-0.5cm}

\section{Semi-analytical analysis}
\label{sec:Analysis}
\vspace{-0.2cm}
In this section, we extend different concepts introduced in the previous sections by means of semi-analytical methods.\\
We consider two systems: the Star/Planet/Satellite (E.g. Sun-Earth-Moon) system and Galcatic-centre/Sun/Planet system. For both systems, we derive the limits of the sphere of influence and the aforementioned spheres of activity. 
\vspace{-0.5cm}

\subsection{Gravitational Spheres of a celestial object}
\vspace{-0.2cm}
The parameters used in this section are extracted from the online NASA planetary and Sun fact sheets.
\vspace{-0.35cm}

\subsubsection{The Sun-Earth-Moon system}
\label{sect:SunEarthMoon}
\vspace{-0.2cm}
We consider here a Star/Planet/Satellite system, that mostly resembles the Sun-Earth-Moon system. The parameters used for this system are summarised in Table \ref{tab:System1}.
\begin{table}
\begin{center}
\begin{tabular}{lcc}
\hline
    & Earth & The Moon \\
\hline   %
  \textbf{Orbital properties}& & \\
  semi-major axis $a$ (AU) & 1.00000 & 0.00257 \\
  eccentricity $e$ & 0.000\,\footnotemark &  0.055 \\
\hline	%
  \textbf{Physical properties}& & \\
 Mass (kg) & $5.97\times 10^{24}$ & not required \\  
 Bulk density (kg/m$^3$) & 5\,514 & 3\,340 \\
 Radius (km) & 6\,378\,& not required\\
\hline   %
\end{tabular}
\caption{\label{tab:System1} The parameters used for the Sun-Earth-Moon system to investigate the ratio of the accelerations as well as the limits of the gravitational spheres (source: online NASA planetary fact sheet).}%
\end{center}
\end{table}
\footnotetext{Though Earth's orbit is slightly eccentric ($e=0.017$), and as we are not searching for an exact solution, we have taken a nil eccentricity for the sake of simplicity.}

For each point along the Moon's orbit, we compute the associated accelerations given by Equations (\ref{eqn:PerturbationR}), (\ref{eqn:PerturbationR1ouF}), and (\ref{eqn:PerturbationF1}). We then evaluate the ratios of Solar accelerations $\nicefrac{F}{R}$ (central/perturbing) and the ratio of the planetary accelerations $\nicefrac{F_1}{R_1}$ (central/perturbing).

\begin{figure}
  \centering
 \includegraphics[width=8.8cm]{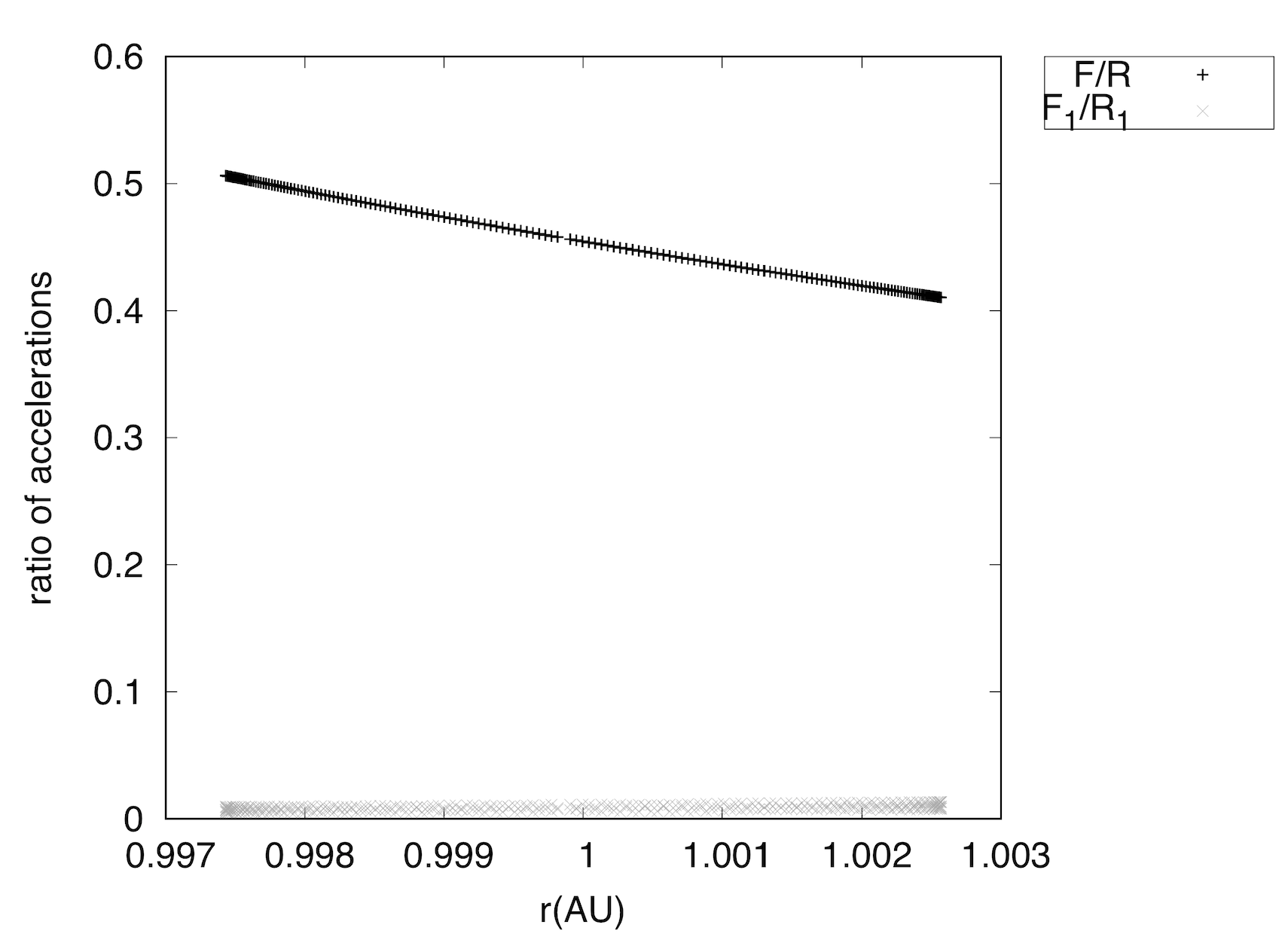}\\ %
 \vspace{-0.2cm}
   \caption{\label{fig:ratioAccelerations} For each point along the lunar's orbit, we evaluate the ratios of Solar accelerations $\nicefrac{F}{R}$ (in black), and the ratio of the planetary acceleration $\nicefrac{F_1}{R_1}$ (in grey) as a function of $r$, the satellite's heliocentric distance.}
\end{figure}

Fig. \ref{fig:ratioAccelerations} shows the ratios $\nicefrac{F}{R}$ and $\nicefrac{F_1}{R_1}$ as a function of the heliocentric distance of the satellite $r$ (cf. Fig. \ref{fig:3bp-rep} for the geometry of the problem). We clearly see that, at all times along the lunar's orbit, the inequation (\ref{ineqn:ActSphere}) is satisfied.
We give in Table \ref{tab:SpheresRadiiEarth0} the values of the Roche limit as well as the Chebotarev, Laplace, Hill, and $\Delta_2$ radii for planet Earth.

\begin{table}
\begin{center}
\begin{tabular}{SlScSc}
\hline
\multirow{2}{*}{  Quantity}	&\multicolumn{2}{c}{  value} \\
    	& min (i.e. $\cos^2\varphi=1$) & max (i.e. $\cos^2\varphi=0$)  \\
 \hline%
    Roche limit & \multicolumn{2}{c}{$ \sim 18\,090\,$\,km ($\sim 0.0001208$\, AU)}\\
  $R_{_\text{Hill}}$ & \multicolumn{2}{c}{0.0100 AU }\\   
  \hline    
\multicolumn{3}{c}{ \textbf{Assuming Earth on a circular orbit ($e=0$)} }	\\	%
    $R_{_\text{Ch}}$  &  \multicolumn{2}{c}{0.00173 AU }\\
 $\Delta_1$ &  $R_{_\text{Laplace}}=$0.00618 AU & 0.00815 AU \\ 
  $\Delta_{_2}$ & 0.0144 AU & 0.0182 AU \\  
  \hline 
\end{tabular}
\caption{\label{tab:SpheresRadiiEarth0} Values of Chebotarev, Hill, Laplace, $\Delta_2$, and Roche radii for Earth (cf. associated parameters in Table \ref{tab:System1}), given the Sun's mass of $\text{M}_\odot=1.9890\times10^{30}\,$kg.}
\end{center}
\end{table}

In Table \ref{tab:SpheresRadiiEarth0}, we give the value of these different radii for planet Earth which we consider on a circular orbit ($e=0$).
The Roche limit depends solely on the physical properties of the planet and the satellite (cf. Section \ref{sect:ROcheLimit}). 

 Unlike the Hill Radius $R_{_\text{Hill}}$ which depends only on the semi-major axis $a$ and the reduced mass $\bar{m}$, the other radii ($R_{_\text{Laplace}}$ and $R_{_\text{Ch}}$) depend on the reduced mass and the heliocentric distance $r_1$. In addition to that, the radii $\Delta_1$ and $\Delta_2$ are very much related to the geometry of the problem (cf. previous section) as they also depend on the angle ($\varphi$).

When appropriate, we have evaluated these radii at their minima and maxima (with respect to $\cos^2\varphi$). The Laplace radius is $R_{_\text{Laplace}}=\Delta_1\Big|_{\cos^2\varphi=0}$.\\
Furthermore, one can see from Table \ref{tab:SpheresRadiiEarth0} that $\Delta_2\Big|_{\cos^2\varphi=0}=1.44\,R_{_\text{Hill}}$ (in agreement with the expected results of Section \ref{sect:discussions}).\\
In what follows, we will always compute the radii at $\cos^2\varphi=0$.

\begin{table}
\begin{center}
\begin{tabular}{SlScSc}
\hline
\multirow{3}{*}{  Quantity}	&\multicolumn{2}{c}{  value} \\
    	& at perihelion & at aphelion \\
    	& $r_1=a\,(1-e)$ & $r_1=a\,(1+e)$ \\
 \hline  
    $R_{_\text{Ch}}$ & 0.00170  & 0.00176 \\
 $R_{_\text{Laplace}}=\Delta_1\Big|_{\cos^2\varphi=0}$ & 0.00607  & 0.00628 \\ 
  $\Delta_{_2}$ & 0.0142 & 0.0146 \\  
  \hline 
\end{tabular}
\caption{\label{tab:SpheresRadiiEarth} We assume Earth on an elliptical orbit ($e=0.017$). This table gives values of Chebotarev, Laplace, and $\Delta_2$ radii for the Sun-Earth-Moon system (cf. Table \ref{tab:System1}), given the Sun's mass of $1.9890\times10^{30}\,$kg, Earth eccentricity $e=0.017$.}
\end{center}
\end{table}

Table \ref{tab:SpheresRadiiEarth}, gives the values of $R_{_\text{Ch}}$, $R_{_\text{Laplace}}$, and $\Delta_{_2}$ for Earth considering the real eccentricity of Earth's orbit ($e=0.017$).
\vspace{-0,2cm}

\subsubsection{The Galactic stellar disc-Sun-Planet system}
Table \ref{tab:SpheresRadiiSun} gives these physical parameters for the Sun which we extracted from the online NASA Sun fact sheet. We consider the Sun's orbit to be circular with a semi-major axis of $\sim 25\,000\,$light-years (common knowledge), with $e=0$.

\begin{table}
\begin{center}
\begin{tabular}{lcc}
\hline
    \textbf{Physical properties of the Sun}   &  value \\
\hline   
 Mass &  $1,9885\times 10^{30}\,\text{kg}=1\,\text{M}_\odot$\\
 Bulk density (kg/m$^3$) & 1\,408 \\
 Radius (km) & 695\,700  \\
\hline   
\end{tabular}
\caption{\label{tab:System2} The physical parameters for the Sun used in the study of the \textit{Galactic stellar disc-Sun-Planet system} to investigate the ratio of the accelerations as well as the limits of the Sun's gravitational spheres (source: online NASA Sun fact sheet).}%
\end{center}
\end{table}

We apply the same reasoning as in Section \ref{sect:SunEarthMoon}, for the Galactic stellar disc-Sun-Earth system. The mass of the Galactic stellar disc is taken from \citep{2016arXiv161207781L}, $M=2.32\times 10^{11}\text{M}_\odot$.

In Table \ref{tab:SpheresRadiiSun}, we give the radii of the different spheres of activity and influence for the Sun when studying the \textit{Galactic stellar disc-Sun-Earth system}.

\begin{table}
\begin{center}
\begin{tabular}{SlScSc}
\hline
\multirow{2}{*}{  Quantity}	&\multicolumn{2}{c}{  value} \\
    	& value (in {ly}) & value (in \text{pc})  \\
 \hline 
    $R_{_\text{Ch}}$  & 0.0519  &  0.0159 \\
 $\Delta_1\Big|_{\cos^2\varphi=0} = R_{_\text{Laplace}}$ &  0.711 & 0.218 \\ 
   $R_{_\text{Hill}}$ & 2.82 & 0.865  \\   
 $\Delta_{_2}\Big|_{\cos^2\varphi=0}$ & 5.13 &1.57 \\  
  \hline 
\end{tabular}
\caption{\label{tab:SpheresRadiiSun} We assume the Sun on a circular orbit ($e=0$). This table gives
values of Chebotarev, Laplace, Hill, and $\Delta_2$ radii for the Galactic stellar disc-Sun-Planet system (cf. Table \ref{tab:System2}).}%
\end{center}
\end{table}
\vspace{-0.30cm}

\subsection{Qualitative analysis of the problem}\label{sect:Analysis} 
\begin{figure}
  \centering
 \includegraphics[width=9.0cm]{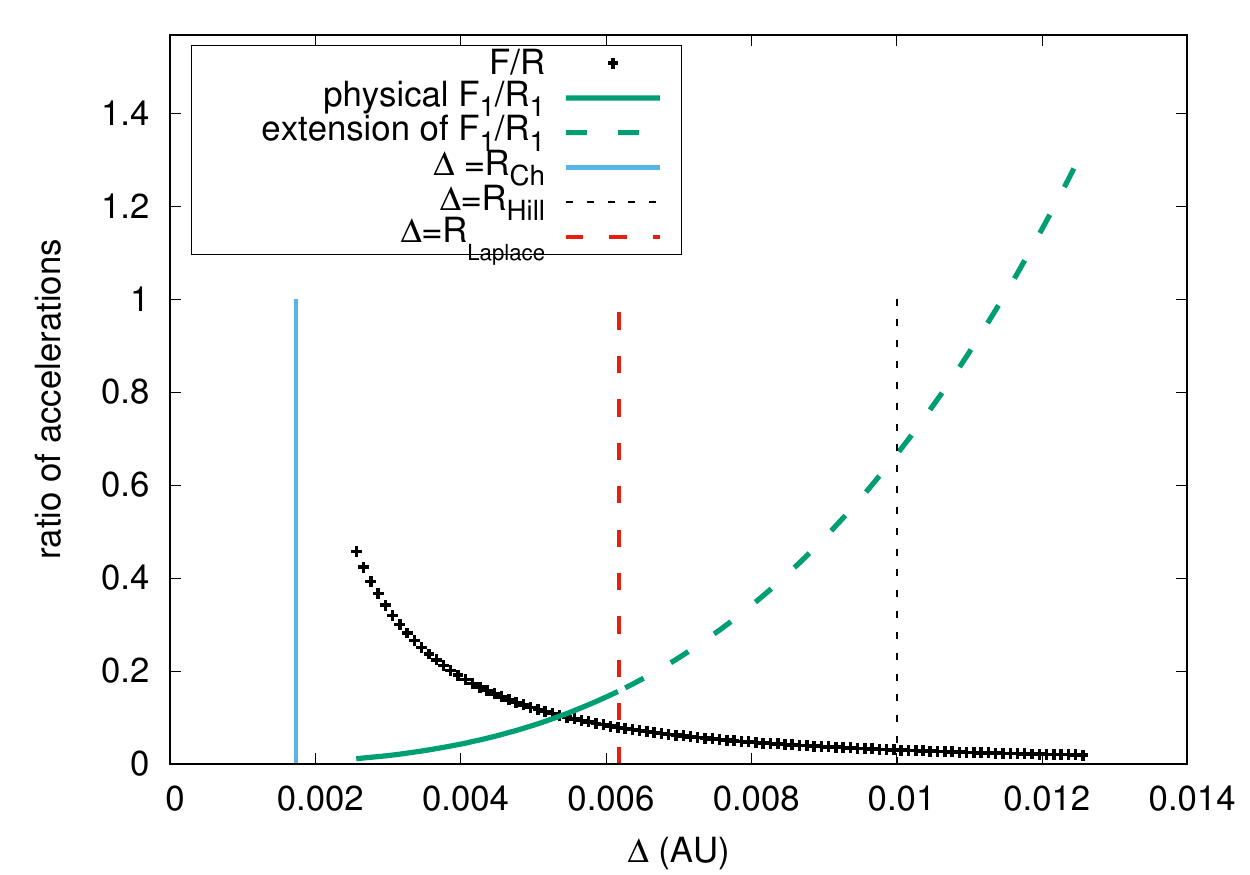}\\ %
   \caption{\label{fig:ExploreAccelerations}Plot of the ratios $\nicefrac{F}{R}$ and $\nicefrac{F_1}{R_1}$ as a function of the geocentric distance of the satellite ($\Delta$) associated with $\cos^2\varphi=0$ and the parameters given in Table \ref{tab:System1}. We indicate relevant quantities such as the Chebotarev radius, the Laplace radius, and the Hill radius (see below for in depth discussion).}
\end{figure}

Fig. \ref{fig:ExploreAccelerations} allows us to fully understand the problem by taking a closer look at the four accelerations in place (cf. Section \ref{sect:SOI}), as well as the limits of the different gravitational spheres of influence (Sphere of activity, Chebotarev, Laplace, and Hill spheres).\\ Fig. \ref{fig:ExploreAccelerations} therefore allows a better understanding of the differences between the Chebotarev, Laplace, and Hill Spheres.  

For a planet Earth on a perfectly circular orbit (e=0.000) at 1.000 AU from the Sun (cf. Table \ref{tab:System1}) and at $\cos^2\varphi=0$, we have plotted the ratios $\nicefrac{F}{R}$ and $\nicefrac{F_1}{R_1}$ as a function of the geocentric distance of the satellite $\Delta$. This plot is only a snapshot for a given geometric arrangement ($\cos^2\varphi=0$) and does not represent the full extent of the problem. It is meant to give us a general view of the problem. In particular, it shows:
\vspace{-0.2cm}
\begin{enumerate}
\item in black crosses the ratio of the of Solar accelerations $\nicefrac{F}{R}$ (central/perturbing). From the expression of $F$ and $R$ given by Equations (\ref{eqn:PerturbationR}) and (\ref{eqn:PerturbationR1ouF}), respectively; we see that  
\begin{equation}
\frac{F}{R} \propto \frac{1}{\Delta^2}\quad ,
\end{equation}
\item in a green continuous bold line the ratio of the planetary acceleration $\nicefrac{F_1}{R_1}$ (central/perturbing). From the expression of $F$ and $R$ given by Equations (\ref{eqn:PerturbationR1ouF}) and (\ref{eqn:PerturbationF1}), respectively; we see that 
\begin{equation}
\frac{F_1}{R_1} \propto \Delta^3 \quad . 
\end{equation}
Furthermore, as a consequence of Equation (\ref{eqn:ratioF1R1}), this ratio can only exist up to this limiting point. The extension to the mathematical function beyond the physical limits is shown in green dashed lines.

\item a light blue vertical line given by the equation $\Delta=R_{_\text{Ch}}$, at which $R=R_1$, as defined in Section \ref{sect:CheboSphere}.
\item a red vertical dashed line given by the equation $\Delta=R_{_\text{Laplace}}$
\item a black vertical dotted line given by the equation $\Delta=R_{_\text{Hill}}$, where $\frac{F_1}{R_1}=\nicefrac{2}{3}$.
\end{enumerate}

\subsection{Further discussion of Fig. \ref{fig:ExploreAccelerations}}
We discuss here the physical meaning of each curve plotted in Fig. \ref{fig:ExploreAccelerations}. The purpose of this paper is to distinguish between the different regimes; a detailed numerical analysis of these regimes is beyond the scope of this paper and will be the subject of future publications.\\

We distinguish three main regions:
\begin{itemize}
\item[$\bullet$] The Chebotarev regime The region where $\Delta \le R_{_\text{Ch}}$:
\begin{itemize}
\item This is the region delimiting the Chebotarev sphere. 
\end{itemize}
\item[$\bullet$] The Laplace regime: The region where $R_{_\text{Ch}} < \Delta < R_{_\text{Laplace}}$:
\begin{itemize}
\item Within this region, we distinguish two different sub-regimes that are separated by the critical value of $\Delta=\Delta_\text{crit}$ at which the two plots of $\nicefrac{F}{R}$ and $\nicefrac{F_1}{R_1}$ intersect.
\begin{itemize}
\item To the left of this point (i.e $\Delta < \Delta_\text{crit}$), we can clearly see that $\frac{F}{R}> \frac{F_1}{R_1}$.
\item To the right of this point (i.e $\Delta < \Delta_\text{crit}$),  $\frac{F}{R}< \frac{F_1}{R_1}$.
\end{itemize}
\end{itemize}
\item[$\bullet$] The Hill regime: The region where $R_{_\text{Laplace}} < \Delta < R_{_\text{Hill}}$:
\begin{itemize}		
\item We are still within the Hill sphere, but we are no longer within the Laplace sphere.	
\end{itemize}
\end{itemize}
In the region where $\Delta > R_{_\text{Hill}}$, no stable orbit around the planet can exist at these geocentric distances.

\section{Conclusion}
In this paper, we have revisited an old, yet still puzzling, question in Classical Mechanics, that of defining the limits of a celestial object's gravitational sphere of influence. As aforementioned, Laplace and his successors such as Hill have investigated the question and provided different answers. 

The Hill sphere would be searching for the region of stable orbits around that planet, whereas the Laplace sphere delimits the region in space of gravitational influence of a planet to deflect a passing comet \cite{Laplace1878} or an asteroid from its initial Keplerian orbit for example.

In our work, we have undertaken an approach aligned with that of \cite{Laplace1878} and \cite{1964SvA.....7..618C}. We have taken a closer look into the satellite's equations of motion and the accelerations that are involved. For a Star-Planet-Satellite system, extending on the work of \citep{1964SvA.....7..618C}, we take a close look at the accelerations involved in the satellite's motion: Solar accelerations (central and perturbing) and planetary accelerations (central and perturbing).

This paper relies on the use of analytical and semi-analytical methods, thus defining a general theoretical framework for future works. Our most significant results can be summarised as follows:
\vspace{-0.15cm}
\begin{itemize}
\item The expressions in Table \ref{tab:summary} summarise the mathematical and/or physical volumes of influence.
\item In Fig. \ref{fig:ExploreAccelerations}, we show an example of how the ratios of the Solar accelerations (central/perturbing) and planetary accelerations (central/perturbing) vary with the geocentric distance of the satellite.
\item Furthermore, we have identified different dynamical regimes which give new directions for future studies to investigate.
\end{itemize}

As a direct application to this paper, we will be investigating the existence of planetary spacing patterns in exo-planetary systems data. 
Over the years, several studies have searched for a Titius-Bode like law.  \citep{1997A&A...322.1018N} and \citep{1998Icar..135..549H}, for example, have calibrated their Titius-Bode like law by setting an an outer cut-off border chosen in an ad hoc way. 

This paper allows us to properly determine the physical outer border of the solar system. Moreover, it has allowed us to identify different regimes (Section \ref{sec:Analysis}). A more in-depth dynamical and numerical analysis of these regimes shall be the subject of future studies.

\section*{Acknowledgements}
\vspace{-0.2cm}
 We thank the reviewer, Professor William \textsc{Newman}, for his thorough revision of our paper and his constructive comments that significantly improve the quality of this work.
 
It is a pleasure to thank Stephanie \textsc{Fearon} and Jon \textsc{Marchant} for their thorough and constructive review of our paper.



\appendix
\section*{Appendix - The case of the giant planets\label{Sect:AppendixGiant}}
\vspace{-0,2cm}
Considering the 3-body problem Sun-Planet-planetesimal\footnote{Not accounting for mutual interactions between the satellites.}, we investigate here the radii of the aforementioned spheres in the case of the four Giant Planets. 
Table \ref{tab:GiantPlanetsParameters} gives for each planet the semi-major axis and eccentricity of the planet's orbit as well as the planet's mass and diameter.

\begin{table}
\begin{center}
\begin{tabular}{lcccc}
\hline
    & $a$ (AU) & $e$ & $D$ & $\nicefrac{1}{\bar{m}}$ \\
\hline   
Jupiter & \,\,5.2026 & 0.048 & 142\,984 & 1\,047.35 \\
Saturn & \,\,9.5549 & 0.056 & 120\,536 & 3\,497.90 \\
Uranus & 19.2184 & 0.046 & \,\,51\,118& 22\,902.94 \\
Neptune & 30.1104 & 0.009 & 49\,528 & 19\,412.24 \\
\hline   
\end{tabular}
\caption{\label{tab:GiantPlanetsParameters} The parameters used for the Giant planets. The semi-major axes ($a$), eccentricities ($e$), and inverse reduced masses ($\nicefrac{1}{\bar{m}}$) are retrieved from \citep{2019Temps}; whereas the diameters are retrieved from the online NASA planetary fact sheets.}%
\end{center}
\end{table}
Table \ref{tab:SpheresGiantPlanets} gives for each one of the Giant Planets: the Roche limit, and the radii $R_{_\text{Ch}}$, $R_{_\text{Laplace}}$, $R_{_\text{Hill}}$, and $\Delta_{2_{\text{inf}}}$.

\begin{table*}
\begin{center}
\begin{tabular}{cScScScScSc}
\hline
\backslashbox{Planet}{Quantity}
& Roche limit (km) &  $R_{_\text{Ch}}$\,(AU)  & $R_{_\text{Laplace}}$\,(AU) & $R_{_\text{Hill}}$\,(AU)   & $\Delta_{2_{\text{inf}}}$\,(AU) \\
  &  & min\,/\,max &min\,/\,max & min\,/\,max &  min\,/\,max \\
\hline 
 Jupiter    & 175\,870 & 0.153\,/\,0.168 & 0.307\,/\,0.338 & 0.338\,/\,0.372 & 0.387\,/\,0.426  \\	
Saturn  & 148\,259 & 0.153\,/\,0.171 & 0.345\,/\,0.386  & 0.412\,/\,0.461 &0.472\,/\,0.528  \\	
 Uranus   & \,\,62\,875 &0.121\,/\,0.133 & 0.331\,/\,0.362 & 0.448\,/\,0.491 & 0.512\,/\,0.562  \\	
 Neptune & \,\,60\,919 & 0.214\,/\,0.218 & 0.575\,/\,0.585  & 0.769\,/\,0.784 & 0.881\,/\,0.897  \\	
\hline
\end{tabular}
\caption{\label{tab:SpheresGiantPlanets} Table summarising for each of the Giant Planets the values of Roche, Chebotarev, Hill, Laplace, and $\Delta_2$ radii (cf. associated parameters for each planet are those in Table \ref{tab:GiantPlanetsParameters}, given the Sun's mass of $\text{M}_\odot=1.9890\times10^{30}\,$kg)}{\textbf{Note:} The Roche limit values given here are computed using the simplified expression, where the Roche limit is 2.46\,planetary radii.} 
\end{center}
\end{table*}

We take a closer look at these results and analyse the distribution of the rings and satellites:
\vspace{-0.2cm}
\begin{itemize}
\item The Jovian System (rings and 79 satellites listed in \citep{2019Temps}\footnote{BDL and IMCCE stand for Institut de M\'ecanique C{\'e}leste st de Calcul des Eph{\'e}m{\'e}rides and Bureau des Longitudes, respectively. These two institutions are both authors and editors of this book.}): while the rings and the satellites Metis and Adrastea lie inside the Roche limit area, the four Galilean satellites and 12 other satellites are found to be at all times between the Roche limit and $R_{_\text{Ch}}$ (in the Chebotarev regime). The remaining 61 satellites are found between the Roche limit and $R_{_\text{Laplace}}$ (with orbits crossing both Chebotarev and Laplace regimes). The satellite (S/2003 J2) is at all times outside of the Chebotarev radius, in the Laplace regime.
\item The Saturnian System (rings and 57 satellites listed in \citep{2019Temps}): Within the Roche limit we find the A, B, and C rings as well as the 6 satellites: Pan, Daphnis, Atlas, Prometheus, Pandora, and S\slash2009 S1. 
The remaining satellites are found within the Laplace sphere, with 17 of them alternating between the Chebotarev and Laplace regimes.
\item The Uranian System (rings and 27 satellites listed in \citep{2019Temps}): The rings as well as three satellites (Cordelia, Ophelie, and Bianca) are found  inside the Roche limit. When 19 satellites are found to orbit at all time in the Cheboratev sphere, the remaining five (Sycorax, Margaret, Prospero, Setebos, and Ferdinand) are found within the Laplace sphere alternating between the Chebotarev and Laplace regimes.
\item The Neptunian System (the rings, the 13 satellites given in \citep{2019Temps}, and the newly discovered satellite, Hippocamp \citep{2019Natur.566..350S}): This system is unique, as it not only presents rings within and outside the Roche limit, but it also presents a structure of arcs (incomplete rings, see \citep{2014A&A...563A.133R} and references therein). Inside the Roche limit, we find the Galle, Le Verrier, Lassell, and Arago rings, as well as the three satellites: Naiad, Thalassa, and Despina. In the Chebotarev regime, we find the Arcs, the Adams ring, as well as the satellites Galatea, Larissa, Hippocamp, Proteus, Triton, Nereid, Halimede, and Sao. On the other hand, the three external satellites Laomedeia, Psamathe, and Neso are found in the Laplace regime.
\end{itemize}

Any detailed numerical analysis of these regimes and the implications of these systems is beyond the scope of this paper and shall be the subject of future studies.


\bsp	
\label{lastpage}
\end{document}